\documentstyle[amssymb,aps,preprint]{revtex}

\begin{document}
\draft
\title{Toward the multiphoton parametric oscillators}
\author{G.Yu.Kryuchkyan and N.T.Muradyan}
\address{Institute for Physical Research, National Academy of Sciences,\\
Ashtarak-2, 378410, Armenia}
\maketitle

\begin{abstract}
We propose novel types of parametric oscillators generating both
three-photon and four-photon bright light which are accessible for an
experiment. The devices are based on the cascaded down-conversion processes
and consist of second-order media inserted in two-resonant mode cavity.
Discussion of dissipation and quantum features of the system are performed
by the quantum-jump simulation method and concerns to the Wigner functions.
The phase-space multistabilities and critical threshold behavior of three-
and four-photon subharmonics are obtained.
\end{abstract}

\pacs{PACS number(s): 42.50.Lc, 42.65.S, 42.50.Dv, 42.65.Yj  }

\section{The problem}

\label{s1}

Optical parametric oscillators (OPO's) based on processes of down-conversion
in a cavity have proved to be efficient sources of light with a range of
unique properties. Among them it should be noted the nonclassical nature of
light, including quantum correlations, sub-Poissonian statistics and a
larger amount of squeezing \cite{numb1}. OPO provides a method for producing
tunable radiation from a coherent fixed-frequency pump source in the
intracavity down-conversion, where a pump-photon splits into a pair of
subharmonic photons. So far OPO has only been realized experimentally for
two-photon down-conversion in nonlinear media with second-order $\chi ^{(2)}$
susceptibility. Experimental and theoretical results can be found in \cite
{numb2}.

In this letter we propose a new conception for the realization of
multiphoton parametric oscillators using the idea of composite multiphoton
interaction. Two schemes of OPO based accordingly on a three-photon and
four-photon intracavity down-conversion are presented.

One of the principal motivations for the investigation of multiphoton OPO
concerns to a production of nonclassical states of light with the novel
features. Other motivation is their possible exploitation for quantum
information technologies, because sources offering a great variety of
multiparticle entangled states, first of all meaning photons, are required
for the implementation of many quantum communication processes such as
direct transmission of the state, quantum teleportation, quantum
cryptography, and entanglement swapping \cite{numb3}.

It is known that two-photon down-conversion so far is the most standard
source of pair of polarization-entangled photons which is relevant to many
problems of quantum information (see, previous experiments on photon
splitting in \cite{numb4}, and for polarization-entangled photon pairs \cite
{numb5}). Although two-particle entanglement has long been demonstrated
experimentally the realization of entanglement between three or more
particles has been considered as a difficult problem. Among the possible
sources of three-photon states (so-called Greenberger-Horne-Zeilinger states 
\cite{numb6}) we mention the process of degenerate three-photon
down-conversion $\omega \rightarrow \omega /3+\omega /3+\omega /3$
\thinspace \thinspace in a $\chi ^{(3)}$ medium. This process has been
studied for running waves \cite{numb7}. Note, that most recently the
polarization entanglement for three spatially separated photons has been
observed \cite{numb8}. Three-photon down-conversion in an optical cavity has
also been proposed and investigated in \cite{numb9}, but too weak
interaction between photons in $\chi ^{(3)}$ medium makes experimental
realization of this scheme a difficult problem. As regards to a four-photon
splitting $\omega \rightarrow \omega /4+\omega /4+\omega /4+\omega /4$
\thinspace \thinspace in a $\chi ^{(4)}$ medium, as we know, this process
was not considered yet.

This letter reports the analysis of two schemes of multiphoton OPO in a
double resonant-mode cavity involving two crystals of second-order
susceptibilities. They are the following:

\subsection{ Four-photon splitting in cascaded down-conversion}

The scheme proposed contains two sequential nonlinear processes of
degenerate two-photon splitting in $\chi ^{(2)}$ media taking place inside
the same cavity. This scheme for the special configuration when the coupling
in- and out-fields occurs at one of the ring cavity mirrors is shown in
Fig.1. The pump frequency $\omega $ is converted to the subharmonic
frequency $\omega _{2}=\omega /2$ and then into the frequency $\omega
_{1}=\omega /4$ in the cascaded parametric processes $\omega \rightarrow
\omega _{2}+\omega _{2}$ and $\omega _{2}\rightarrow \omega _{1}+\omega
_{1}. $ We assume that the each of these processes takes place effectively
only in the definite nonlinear crystal. The cavity is also assumed
transparent at the pump frequency. So, we neglect the pump depletion effects
considering the amplitude of driving field as a classical constant. We model
the combination of these processes by the interaction Hamiltonian: 
\begin{equation}
H_{1}=i\hbar \chi _{1}(Ea_{2}^{\dagger 2}-E^{\ast }a_{2}^{2})+i\hbar
k_{1}(a_{1}^{\dagger 2}a_{2}-a_{1}^{2}a_{2}^{\dagger }),  \label{ham1}
\end{equation}
where $a_{1}$ and $a_{2}\,$ are the operators of the modes $\omega _{1}$ and 
$\omega _{2},$ $\chi _{1}$ and $k_{1}$ are the nonlinear coupling constants
for the processes $\omega \rightarrow \omega _{2}+\omega _{2}$ and $\omega
_{2}\rightarrow \omega _{1}+\omega _{1}$ respectively which are related to
second-order nonlinear susceptibilities $\chi ^{(2)}$. $E$ is proportional
to the coherent driving field amplitude.

\subsection{Three-photon splitting in cascaded down-conversion}

This system mainly is similar to the previous one. The only difference is
that the pump frequency $\omega $ converts to two different frequencies$\
\omega _{1}=\omega /3$ and $\omega _{2}=2\omega /3$ in the parametric
interaction $\omega =\omega _{1}+\omega _{2}.$ The subharmonic $\omega _{1}$
is also produced in the other process of down conversion $\omega _{2}=\omega
_{1}+\omega _{1}.$ On the whole two subharmonic modes $\omega _{1}=\omega /3$
and $\omega _{2}=2\omega /3$\ are created in the cavity due to the cascaded
processes. In the undepleted pump field configuration this model is
described by the following interaction Hamiltonian: 
\begin{equation}
H_{1}=i\hbar \chi _{2}(Ea_{1}^{\dagger }a_{2}^{\dagger }-E^{\ast
}a_{1}a_{2})+i\hbar k_{2}(a_{1}^{\dagger 2}a_{2}-a_{1}^{2}a_{2}^{\dagger }),
\label{ham2}
\end{equation}
where $a_{1}$ and $a_{2}\,$ are the operators of $\omega _{1}$ and $\omega
_{2}$ subharmonics$,$ and $\chi _{2}$ and $k_{2}$ are the coupling constants
of the processes $\omega \rightarrow \omega _{1}+\omega _{2}$ and $\omega
_{2}\rightarrow \omega _{1}+\omega _{1},$ respectively$.$

We note, that optical schemes involving cascaded nonlinearities in a cavity
are currently of considerable interest and hold promise for applications
including, in particular, frequency tunable sources of light, large
third-order nonlinear effects via cascaded second-order nonlinearities and
squeezing phenomena. The references and experimental observation can be
found in \cite{numb10}. Possible operational regimes of a combined system,
as a rule, essentially differ from those in pure processes in each of
nonlinear media. Some results in this direction are devoted to the cascaded
frequency doubler \cite{numb11}. Concerning our schemes, it is interesting
to note, that the stable stationary above-threshold regime of oscillation is
realized even in undepleted pump regime. The other novelty here is
comparatively low values of the threshold of generation due to the schemes
proposed can be accessible for experiments. Contrary to
polarization-entangled state generation in two- and three-photon devices 
\cite{numb5,numb8} the basic multiphoton states of our systems are neither
based on polarization nor on momentum, but on phase-space entanglement of
modes.

\section{Four-photon OPO}

The system of interest is dissipative because the subharmonic modes suffer
losses due to partially transmission of light through the mirrors of the
cavity. So that the reduced density operator $\rho $ of both modes obeys a
master equation 
\begin{equation}
\frac{\partial \rho }{\partial t}=\frac{1}{i\hbar }\left[ H_{1},\rho \right]
+\sum_{i=1,2}\gamma _{i}(2a_{i}\rho a_{i}^{\dagger }-a_{i}^{\dagger
}a_{i}\rho -\rho a_{i}^{\dagger }a_{i}),  \label{master1}
\end{equation}
where $\gamma _{1}$\ and $\gamma _{2}$\ \ are the cavity damping rates for
the modes $\omega _{1}$ and $\omega _{2}.$

To proceed with further analysis we present now the semiclassical
steady-state solutions and stability properties of the system proposed. In
the classical limit the system may be described by the complex-field
amplitudes $\alpha _{1}$ and $\alpha _{2}$ of the resonant modes $\omega
_{1} $ and $\omega _{2}$ whose real and imaginary parts respectively
represent the dimensionless position and momentum. They obey the equation of
motion: 
\begin{eqnarray}
\frac{\partial \alpha _{1}}{\partial t} &=&-\gamma _{1}\alpha
_{1}+2k_{1}\alpha _{2}\alpha _{1}^{\ast },  \nonumber \\
\frac{\partial \alpha _{2}}{\partial t} &=&-\gamma _{2}\alpha _{2}+2\chi
_{1}\alpha _{2}^{\ast }E-k_{1}\alpha _{1}^{2}.  \label{lang1}
\end{eqnarray}
Solving these equations for the steady-state regime and carrying out the
standard linearized stability analysis one can arrive at the following
results.

The trivial zero-amplitude solution $\alpha _{1}=\alpha _{2}=0$\ is stable
in the region $E<E_{th},$ $E_{th}=\gamma _{2}/2\chi _{1},$ with $E_{th}$
being the threshold value of the pump field amplitude, and it describes the
below-threshold regime of oscillation of both subharmonics. The stable
above-threshold solutions, for $E>E_{th}$, expressed in terms of photon
numbers $n_{j}$ and phases $\varphi _{j}$ of subharmonic modes ($\alpha
_{j}=\exp (i\varphi _{j})\sqrt{n_{j}},\,\ j=1,2$) are calculated as: 
\begin{eqnarray}
n_{1} &=&\gamma _{1}\gamma _{2}(E/E_{th}-1)/2k_{1}^{2},\,\ n_{2}=\gamma
_{1}^{2}/4k_{1}^{2},  \label{photon1} \\
\varphi _{1} &=&\frac{\Phi }{4}\pm \frac{\pi }{2}m,\,\ \ \varphi _{2}=\frac{%
\Phi }{2}\pm \pi n,\,\ n,m=0,1,2...,  \label{phase1}
\end{eqnarray}
where $\Phi $ is the phase of the driving field.

So, this scheme of multiphoton OPO allows us to reach stable generation of
subharmonics at frequencies $\omega _{1}=\omega /4$ and $\omega _{2}=\omega
/2$\ at $E>E_{th}$ in undepleted pump-field regime. It should be pointed out
that the occurrence of such a regime of oscillation becomes possible just
due to the combination of two kinds of down-conversion processes in a
cavity. One of the peculiarities of cascading dynamics is that the threshold
pump power $P_{th}=c\hbar \omega E_{th}^{2}/2L$ ($L$ is the optical way of
pump field) is only related to the coupling constant $\chi _{1}$ of $\omega
\rightarrow \omega /2+\omega /2$ parametric process. This coupling constant
is proportional to the second-order susceptibility that leads to the
comparatively low value of the threshold. We illustrate this fact for the
system depicted in Fig.1, choosing realistic losses and typical values of $%
\chi _{1}$ and $k_{1}$ (see detailed calculations for a similar cavity
configuration in \cite{numb11}). Assuming $\chi _{1}=$ $k_{1}=3.4\times
10^{4}s^{-1}$\ , $\gamma _{1}=2.4\times 10^{8}s^{-1}$ , $\gamma
_{2}=3.6\times 10^{8}s^{-1}$, we find for the threshold pump power $%
P_{th}=74mW$. For these parameters the power of the cavity output field of $%
\omega /4$ mode increases as $P_{1}^{out}=2\hbar \frac{\omega }{4}\gamma
_{1}n_{1}=3.4\times 10^{-3}(E/E_{th}-1)W$, while the output power of the $%
\omega /2$\ \ subharmonic remains constant $P_{2}^{out}=2\hbar \frac{\omega 
}{2}\gamma _{2}n_{2}=3.4\times 10^{-3}W.$

A very important characteristic of multiphoton OPO concerning phase
information of the generated modes, is displayed in phase-space. As we see
from (\ref{phase1}), the stationary states of \ $\omega /4$\ \ subharmonic
have four-fold symmetry. There are four stable states of $\omega _{1}$ mode
with equal photon numbers, but with four different phases which are: $\frac{%
\Phi }{4},\frac{\Phi }{4}+\frac{\pi }{2},\frac{\Phi }{4}+\pi ,\frac{\Phi }{4}%
+\frac{3\pi }{2}.$ As regards to $\omega /2$\ \ subharmonic mode, it
displays two-fold symmetry in phase-space, i.e. there exist two states with
equal intensities but with different phases $\frac{\Phi }{2},\frac{\Phi }{2}%
+\pi .$

Let us now show the general character of this symmetry in the frame of a
full quantum treatment of interaction. It is easy to verify that the
Hamiltonian (\ref{ham1}) as well as the density operator $\rho $ obeying Eq.(%
\ref{master1}) satisfy the commutation relations$\,$%
\begin{equation}
\left[ H_{1},U\right] =\left[ \rho (t),U\right] =0  \label{sym1}
\end{equation}
with the operator 
\[
U=\exp \left( i\frac{\pi }{2}a_{1}^{\dagger }a_{1}+i\pi a_{2}^{\dagger
}a_{2}\right) . 
\]
For the reduced density operators of each of the modes $\rho
_{1(2)}=Tr_{2(1)}(\rho ),$ (which are constructed by tracing over one of the
modes), relations$\ $(\ref{sym1}) give$\,$%
\begin{equation}
\left[ \rho _{1}(t),U_{1}(\frac{\pi }{2})\right] =\left[ \rho
_{2}(t),U_{2}(\pi )\right] =0,  \label{symr}
\end{equation}
where operators$\ U_{i}(\varphi )=\exp \left( i\varphi a_{i}^{\dagger
}a_{i}\right) ,\,i=1,2$ perform rotation by the angle $\varphi $ around the
origin in phase-space of complex stochastic variables $\alpha _{1}$ and $%
\alpha _{2}$, respectively.

One of the most important conclusions of such symmetries is related to the
Wigner functions $W_{1}$ and $W_{2}$ of the subharmonic modes $\omega _{1}$
and $\omega _{2}$ which provide a large amount of information about the
states of the mode and also provide a pictorial view. To see this we address
to the formula for Wigner function expressed through the density operator.
Using also formulas (\ref{symr}) we arrive to the symmetries of Wigner
functions written through the stochastic mode amplitudes $\alpha _{1}$ and $%
\alpha _{2}$ corresponding the operators $a_{1}$ and $a_{2}.$ In the polar
coordinates $r,\theta $ of the complex phase-space plane $X=\frac{\alpha
+\alpha ^{\ast }}{2}=r\cos \theta ,\,$ $Y=\frac{\alpha -\alpha ^{\ast }}{2i}%
=r\sin \theta $ these symmetries take the forms: 
\begin{equation}
W_{1}(r,\theta +\pi /2)=W_{1}(r,\theta ),\,\ \ W_{2}(r,\theta +\pi
)=W_{2}(r,\theta ).  \label{wig1}
\end{equation}
\ We have obtained this property of Wigner functions using the symmetry of
Hamiltonian (\ref{ham1}) and dissipation terms in master equation (\ref
{master1}), assuming that the initial state of the modes is a vacuum state.
Below, we will present the other confirmation to phase-space symmetry in
quantum trajectories.

\section{Three-photon OPO}

The dynamics of subharmonic modes $\omega _{1}=\omega /3$ and $\omega
_{2}=2\omega /3$ for this system is governed by master equation (\ref
{master1}) involving Hamiltonian (\ref{ham2}) instead of Hamiltonian (\ref
{ham1}). Accordingly, the classical equations of motion for the complex
amplitudes are 
\begin{eqnarray}
\frac{\partial \alpha _{1}}{\partial t} &=&-\gamma _{1}\alpha _{1}+\chi
_{2}\alpha _{2}^{\ast }E+2k_{2}\alpha _{2}\alpha _{1}^{\ast },  \nonumber \\
\frac{\partial \alpha _{2}}{\partial t} &=&-\gamma _{2}\alpha _{2}+\chi
_{2}\alpha _{1}^{\ast }E-k_{2}\alpha _{1}^{2}.  \label{lang2}
\end{eqnarray}

We list below the semiclassical results for stable values of photon numbers
and phases of the modes. In the above-threshold generation regime, at $%
E>E_{th}$, where

\begin{equation}
E_{th}=\frac{2\sqrt{2\gamma _{1}\gamma _{2}}}{3\chi _{2}}  \label{thresh2}
\end{equation}
is the threshold value of E, the results take the form: 
\begin{eqnarray}
n_{1} &=&\frac{\gamma _{1}\gamma _{2}}{18k_{2}^{2}}\left( \varepsilon +3%
\sqrt{\varepsilon ^{2}-1}\right) ^{2},  \nonumber \\
n_{2} &=&\frac{\gamma _{1}^{2}}{36k_{2}^{2}}\left( \frac{\varepsilon +3\sqrt{%
\varepsilon ^{2}-1}}{\varepsilon +\sqrt{\varepsilon ^{2}-1}}\right) ^{2},
\label{upper}
\end{eqnarray}
where $\varepsilon =E/E_{th},$ 
\begin{equation}
\phi _{1}=\frac{\Phi }{3}+\frac{2\pi }{3}n\,\,\,,\,\,\phi _{2}=\frac{2\Phi }{%
3}-\frac{2\pi }{3}n,\;\;n=0,1,2.  \label{upperf}
\end{equation}

In the region $\varepsilon <\frac{3}{2\sqrt{2}}$ the stability condition is
fulfilled only for a zero amplitude steady-state solution $\alpha
_{1}=\alpha _{2}=0,$ i.e. the subharmonic field excitation exists at the
spontaneous noise level. The photon numbers are plotted against the pump
amplitude in Fig.2. They show bistable hysteresis-cycle behavior in a small
domain $1<\varepsilon <\frac{3}{2\sqrt{2}}.$ The dashed parts of the curves
correspond to unstable solutions. As we see, with increasing $E/E_{th}$ \
the photon number of the $\omega /3$\ subharmonic increases as $n_{1}\simeq 
\frac{8\gamma _{1}\gamma _{2}}{9k_{2}^{2}}\varepsilon ^{2}$, while the
photon number of the $2\omega /3$\ subharmonic goes to the saturated value $%
n_{2}=\gamma _{1}^{2}/9k_{2}^{2}.$

The very important peculiarity of the system proposed is that the threshold
value $E_{th}$ depends on the coupling constant $\chi _{2}$ which is related
to the second-order $\chi ^{(2)}$ susceptibility in contrary to the scheme
cited in \cite{numb9}, where the threshold of the driving field amplitude is
determined by $\chi ^{(3)}$ susceptibility. However, the photon numbers of
the generated modes depend on both nonlinear coupling constants $\chi _{2}$
and $k_{2}$. For the realistic parameters: $k_{2}=2.4\times 10^{3}s^{-1},$ $%
\chi _{2}=2.3\times 10^{3}s^{-1},$ $\gamma _{1}=0.9\times 10^{8}s^{-1}$, $%
\gamma _{2}=1.2\times 10^{8}s^{-1}$ parametric oscillations occur above the
threshold pump power $P_{th}=3W$ and the threshold values of cavity-output
field powers are $P_{1}^{th}\approx 4mW,$ $P_{2}^{th}\approx 4mW$. Eqs.(\ref
{upper}),(\ref{upperf}) imply that in the above-bistability regime there
exist three states for each of the modes which have equal intensities $n_{1}$
and $n_{2}$ but different phases: $\frac{\Phi }{3},\,\frac{\Phi }{3}+\frac{%
2\pi }{3},\,\frac{\Phi }{3}+\frac{4\pi }{3}$ and $\frac{2\Phi }{3},\,\frac{%
2\Phi }{3}-\frac{2\pi }{3},\,\frac{2\Phi }{3}-\frac{4\pi }{3},$
respectively. Surprisingly, both the modes $\omega _{1}=\omega /3$\ and $%
\omega _{2}=2\omega /3$ have the identical threefold symmetry which also
follows from the specific three-photon form of Hamiltonian (\ref{ham2}). In
the full quantum treatment of cascaded three-photon down-conversion this
symmetry, in particular, is displayed in the Wigner functions. Following the
method presented above, we have obtained: 
\begin{equation}
W_{i}(r,\theta +2\pi /3)=W_{i}(r,\theta ),\,\ \ i=1,2.
\end{equation}

\section{Quantum simulation: multistabilities and threshold behavior}

The second part of this paper is devoted to the study of quantum-statistical
properties of cascaded multiphoton OPO in a dissipative cavity. Our aim is
to analyze the phase-space properties of the subharmonic modes in a quantum
regime in the presence of quantum noise on the basis of the Wigner functions.

Note that an exact analysis of quantum optical nonlinear systems including
dissipation effects is a difficult problem which was solved for a few single
models (see \cite{numb12} and references therein). In this field of
research, analytical expression for the Wigner function of two-photon OPO
was derived in \cite{numb13}. However the quantum-statistical properties of
cascaded OPO's can not be analyzed using single analytical formulas. Our
approach is based on the quantum-jump simulation method also known as the
state-vector Monte-Carlo method \cite{numb14}. This method considers not the
density matrix but deals with the state vector $\mid \Psi ^{\alpha
}(t)\rangle $ which is a member of the ensemble of state vectors and index $%
\alpha $ indicates the realizations. The procedure adopted in the
quantum-jump simulation for the system Hamiltonians (\ref{ham1}) and (\ref
{ham2}) consists in the following. The evolution of a member of the ensemble
of pure states $\mid \Psi ^{\alpha }(t)\rangle $ over a short time $\delta t$
is governed by the non-Hermitian effective Hamiltonian if there is no
quantum jump 
\begin{equation}
H_{1(2)eff}=H_{1(2)}-i\hbar (\gamma _{1}a_{1}^{\dagger }a_{1}+\gamma
_{2}a_{2}^{\dagger }a_{2}).  \label{effham}
\end{equation}

Such evolution must be completed by the possibility of quantum jumps which
change the state vectors of both modes. On the whole the state transforms in
the following way 
\begin{equation}
\mid \Psi ^{\alpha }(t+\delta t)\rangle \rightarrow \left\{ 
\begin{array}{ll}
\left( 1-\frac{i}{\hbar }\delta tH_{1(2)eff}\right) \mid \Psi ^{\alpha
}(t)\rangle /(1-\delta p)^{1/2}, & \text{with probability }1-\delta p, \\ 
\sqrt{2\gamma _{i}}a_{i}\mid \Psi ^{\alpha }(t)\rangle /(\delta p_{i}/\delta
t)^{1/2}, & \text{with probability \ }\delta p_{i},\,\,i=1,2,
\end{array}
\right.  \label{jumps12}
\end{equation}
where $\delta p=\delta p_{1}+\delta p_{2}.$

We generalize the method presented in \cite{numb15} for the quantum-jump
simulation of the Wigner functions. Details of analogous calculations for
the second harmonic generation can be found in \cite{numb16}.The numerical
simulations are performed in the truncated Fock basis of both subharmonic
modes $\omega _{1}$\ and $\omega _{2}$\ in the regime of strong nonlinear
coupling between the subharmonic modes, i.e. $\chi _{1,2}\lesssim \gamma
_{1},\gamma _{2}$\ and $k_{1,2}\lesssim \gamma _{1},\gamma _{2}$. We note
that the OPO with such extremely large nonlinearities are not realized in
practice and in this part of the Letter we do not intend to give results
close to an experimental situation but discuss the fundamental problems of
multiphoton entangled states.

To be short, we shall give the results for three-photon OPO, where one kind
of phase-space symmetries is realized. The steady-state Wigner functions for 
$\omega _{1}=\omega /3$\ and $\omega _{2}=2\omega /3$ modes are calculated
in the basis of $n=45$ photon number states for each mode. The results may
be written as depending on the dimensionless parameters: $\varepsilon
=E/E_{th},\,\ k_{2}/\gamma _{1},\;\gamma _{2}/\gamma _{1}.$ We calculate the
Wigner functions of $\omega _{1}=\omega /3$ subharmonic averaged over 1000
trajectories in the complex phase--space plane $X,\,Y.$ It is obvious that
below the threshold the Wigner function is single-humped and centered at $%
X=Y=0$. This state indicates appearance of three arms as the consequence of
small entanglement between the modes. Increasing the driving field $E$ we
enter the bistability domain and observe the occurrence of three additional
side-humps. They correspond to the above-threshold steady-states (12), (13)
with equal photon numbers and threefold symmetric phases. We see explicitly
that in this domain the behavior of our system is actually four-stable. With
further increase of $E/E_{th}$ the central hump disappears, while the
side-humps increase and we turn to a manifestly above threshold oscillation
regime when we observe the phase space three-stability. This case is shown
in Fig.3 for the parameters $k_{2}/\gamma _{1}=0.2,\;\gamma _{2}/\gamma
_{1}=0.4,\;\varepsilon =1.59$. Analogous behavior has the Wigner function of 
$2\omega /3$-subharmonic.. The difference between the results of two modes
is displayed in the form of squeezing the humps.

\section{CONCLUSION}

\label{s5}Thus we have shown that the idea of using the cascaded photon
splitting processes in a cavity opens a novel route for the realization of a
multiphoton parametric oscillator. Two devices are proposed which transform
the pump coherent field into two subharmonics at frequencies $\omega /2$ and 
$\omega /4$ (configuration A) and $\omega /3$ and $2\omega /3$
(configuration B). Furthermore, these models of OPO are to be subject to an
experiment. We have been convinced that the pump threshold of each of the
OPO is only expressed through the second-order susceptibility. As a result,
comparatively low values of thresholds have been obtained for typical
parameters. The important features of multiphoton OPO obtained are
phase-space symmetries and multistabilities of subharmonics. In addition, it
is shown that subharmonics ''remember'' the phase of the pump field. Our
numerical quantum Monte-Carlo simulations display phase-space
multistabilities and threshold behavior of subharmonics in the presence of
dissipation.

\bigskip {\bf Acknowledgments}

We acknowledge the helpful discussions with J.Bergou, H.Carmichael, S.Fauve
and L.Manukyan. The work was supported in part by INTAS grant No 97--1672,
and by grant No 00375 awarded by the Armenian Science Foundation.

{\Large Figure Captions}

Fig.1. Principal scheme of the cascaded parametric oscillator. The cavity is
resonant at two frequencies $\omega _{1}=\omega /4$ and $\omega _{2}=\omega
/2$ of the subharmonics. The phase matching condition for the process $%
\omega \rightarrow \omega /2+\omega /2$ is satisfied in the first medium,
and for the process $\omega /2\rightarrow \omega /4+\omega /4$ in the second
medium.

Fig.2. The normalized mean photon numbers $n_{1}/n_{1}^{th}$ (curve1) and $%
n_{2}/n_{2}^{th}$ (curve 2) of $\omega /3-$ and $2\omega /3-$ modes versus
the scaled pump intensity parameter $\varepsilon ^{2}=E^{2}/E_{th}^{2}$. The
threshold values of the photon numbers are $n_{1}^{th}=\frac{\gamma
_{1}\gamma _{2}}{18k_{2}^{2}}$ and $n_{1}^{th}=\frac{\gamma _{1}^{2}}{%
16k_{2}^{2}}$. The dashed parts of the curves 1,2, and zero-amplitude (curve
3) solutions for both modes describe the unstable steady-state solutions.

Fig.3. The steady-state Wigner function of $\omega /3$ mode which
demonstrate the phase-space multistability in the above-threshold range.

\end{document}